\documentstyle [12pt,preprint,eqsecnum,aps]{revtex}
\begin {document}

\draft

\title {One Dimensional Phonon Coupled Electron Tunneling: A Realistic
Model}

\author {E.Pazy and B.Laikhtman}
\address {Racah Institute of Physics, Hebrew University, Jerusalem 91904,
Israel}

\maketitle \begin{abstract}

The transition probability for a one dimensional tunneling electron
coupled to acoustical phonons is calculated, with the Feynman
path-integral method. We considered a realistic
electron phonon interaction (deformation potential, piezoelectric), making
use of slowness of the phonon system compared to electron tunneling.
We show that the problem of the complex non-linear coupling of a tunneling
electron to the zero point fluctuations of a phonon field is equivalent
to that of an electron tunneling through a slow fluctuating spatially
uniform barrier, thus resulting in an increase of the
tunneling probability due to electron coupling with zero-point phonon
oscillations. We calculated also the energy change of the
tunneling electron due to phonon emission.

\end{abstract}

\section { Introduction }
\label {sec:Introduction}
Tunneling, being one of the most remarkable non-classical manifestations
of quantum dynamics, has been largely investigated. One of the major
issues in the field is the understanding of the properties of tunneling
particles coupled to degrees of freedom representing the environment. The
case of a particle or macroscopic degree of freedom tunneling out of a
metastable state in the presence of arbitrary linear dissipation mechanism,
has been extensively studied in a series of papers by Caldeira and
Leggett \cite {Caldeira 81}
and other groups \cite{Chakravarty 82}.
Two state systems
coupled to a dissipative environment have been studied in depth by a large
number of groups \cite{Sethna81}, in the last  two decades.

The problem that we address in the present paper is the tunneling of an
electron with a given energy $E$ across a rectangular barrier where it is
coupled with phonons. We calculate the change of the tunneling probability due
to this coupling and the average energy loss due to phonon emission of
tunneled electrons. In this work we consider only a one-dimensional problem
that can be realized as a quantum wire, or in narrow constrictions.

This problem, finding the transmission probability and spectrum of a tunneling
electron coupled to phonons is of fundamental interest as well as being
potentially of great technological importance. Examples are tunneling through
Josephson junctions \cite{Kampf 87}, bit errors in mesoscopic logic and memory
circuits \cite{Bandara 90}, as well as quantum cascade lasers \cite{Faist 94},
and of course, scanning tunneling microscopy (STM) \cite{Kaiser 93}. There are
even some recent examples of the use of the transmission spectrum to define an
experimental test between quantum mechanics and a class of alternative
theories \cite{Baublitz 95}. Also recent development of nanostructure
technology makes it possible to check experimentally the effect of coupling
with the environment on the dependence of the tunneling probability on the
width and the height of the barrier.

Typically the coupling between the tunneling particle and
the environment is assumed to be linear with respect to
particle coordinate. This assumption may be justified for tunneling of heavy
atoms or macroscopic degrees of freedom but it is not realistic for
electrons. A nonlinear coupling with phonons was considered by Bruinsma and
Bak \cite{Bruinsma 86}. But they considered only the case when the coupling
between the tunneling particle and each of phonon degree of freedom is the
same, which is not realistic in a regular lattice. Due to this
the result of Bruinsma and Bak contains a cut off frequency. In our realistic
treatment of electron-phonon coupling a characteristic frequency
naturally arises in the theory. Since the only length scale in the
problem is the barrier length and from it a typical time scale emerges.

A recent treatment of the problem was given by M. Ueda \cite{Ueda 96},
who used the effective action technique and got analytic results, for some
cases of the spectral density function of the phonon bath. Whereas
Bruinsma and Bak \cite{Bruinsma 86} calculated the
transmission spectrum, we show that the main effect, on the electron transition
probability, is static i.e a static effective lowering of the barrier
by phonons. Such an effect, a static lowering of the barrier due to
coupling of the tunneling electron to zero point fluctuations, was also
discussed in two papers by Ueda and Ando \cite{Ueda 94} although they
studied a different physical problem, the coupling of the tunneling
electron to electromagnetic modes (in their paper this was referred to
as the dynamic limit).

A deformation potential coupling of tunneling electrons with phonons was
also studied by Gelfand et al.\cite{gelfand}. However, they considered only
an infinitesimally thin barrier.

We consider interaction of tunneling electron with acoustic phonons via
realistic coupling mechanisms: deformation potential or piezoelectric.
Coupling with optical phonons is neglected for the following reasons. The
amplitude for zero point fluctuations for optical phonons, for a given
frequency is proportional to one over the square root of the frequency. Since
the optical frequencies are large compared to those of acoustic phonons we
consider their zero point fluctuations as negligible. Also the coupling
constant with optical phonons in III-V materials \cite{re:effective mass} is
not very large giving a further justification for neglecting the effect of
optical phonons. Concerning optical phonon emission and absorption we study
the low temperature case and assume that electron energy and temperature are
much smaller than the energy of the optical phonons so that emission and
absorption can be neglected.

The coupling with acoustic phonons leads to a modulation of the barrier. This
effect exists even at zero temperature when the modulation is controlled by
zero-point phonon vibrations. In other words, as a result of coupling with
phonons the height of the potential barrier is not any more a classical
parameter determined by fabrication of the structure but a quantum mechanical
dynamical variable that is characterized by a wave function. Thus the
complex non-linear problem of a tunneling electron coupled to a phonon
field is transformed to a that of an electron tunneling through an
effective potential barrier where the barrier height fluctuates.

The paper is composed in the following way. In Sec.\ref{sec:general} we
formulate the model. In Sec.\ref{sec:ma} we introduce two main approximations
which we use to facilitate the calculation of the tunneling probability. Both
coupling mechanisms that we consider are nonlinear with respect to the
electron coordinate. This nonlinearity makes the application of generally used
instanton method difficult and to solve the problem we developed an expansion
on some small parameters. First, it appears that frequencies of typical
phonons interacting with a tunneling electron are so small that the phonon
system can be considered slow compared to the electron. Second, we use the WKB
approximation. The slowness of the phonon system allows us in the first
approximation to neglect phonon dynamics and to consider their potential as
time independent that dramatically simplifies the calculation (the calculation
without this approximation appears in the Appendix. The corresponding
calculation of the transmission coefficient are carried out in
Sec.\ref{sec:statapp}. The static approximation, however, does not include
phonon emission and resulted energy loss. To calculate it a more general
approach is necessary. In Sec.\ref{sec:physical results} again making use of
the slowness of the phonon system we go beyond the static approximation and
calculate the average energy loss of tunneled electrons.

\section {general formulation of the problem}
\label{sec:general}
\subsection{The model}
\label{sb:model}

We treat the problem of a one dimensional electron coupled to acoustic phonons
while tunneling through a rectangular barrier, of length L. The potentials
restricting the electron's movement do not effect phonons, which move freely
through the bulk and thus are treated as three dimensional. Being interested
in the effect of the zero point fluctuations of the phonon field on the
electron tunneling probability we consider the system at zero temperature.
Practically it means that temperature is low enough and this condition as well
as the finite temperature case are discussed in the Appendix. Coupling the
electron to the phonon field transforms the problem from a low dimensional
quantum problem to a field theory problem. A convenient way to approach such a
field theory problem is via the formalism of path integrals \cite{Hibbs65}.

The first stage in the calculation of the transition probability for the
tunneling process is the calculation of the retarded Green function. We
express the energy dependent retarded Green function as the Fourier transform
of the time dependent propagator,
\begin{eqnarray}
\label{eq:prop}
& & K (x_{b}, u_{{\bf q} f} ,E \mid  x_{a}, u_{{\bf q} i})  =
\int_{0}^{\infty} K \left ( x_{b}, u_{{\bf q} f},T \mid x_{a},
u_{{\bf q} i} \right )
\exp{ \left ( \frac{\imath E \cdot T}{\hbar}\right ) } dT \ ,
\end{eqnarray}
that can be written as the combined path integral of the electron and phonons:
\begin{eqnarray}
\label{eq:propagator}
K \left( x_{b}, u_{{\bf q} f},T \mid x_{a}, u_{{\bf q} i} \right) & = &
        \int {\cal D}x \int {\cal D}u_{\bf q} \
        \exp    \left\{ \frac{\imath}{\hbar}
                \left[
        {\cal S}_{el}(x) + {\cal S}_{ph}(u_{\bf q}) + {\cal S}_{int}(u_{\bf q},x)
                \right]
                \right\} ,
\end{eqnarray}
where $x$ is the electron coordinate, $u_{\bf q} $ are the Fourier
coefficients of $u(x,t)$, which is the $x$ component of the displacement
vector and $x_{a}, u_{{\bf q} i}$ ; $x_{b}, u_{{\bf q} f}$ are the initial and
final coordinates of the electron and phonons respectively. The limits over
the energy Fourier transform result from the fact we are calculating the
retarded Green function. $ {\cal S}_{el} $ is the action of the electron in
the absence of phonons, given by:

\begin{equation}
{\cal S}_{el} = \int_{0}^{T}  dt \left ( \frac{m}{2} {\dot x}^2 -
V \right),
\label{eq:Selectron}
\end{equation}
where, $m$, is the effective electron mass, and $V$ is the constant potential
height of the rectangular barrier. ${\cal S}_{ph}$ is the action of the
uncoupled phonon field:

\begin{equation}
{\cal S}_{ph} = \int_{0}^{T}  dt \sum_{q}  \left (
\frac{\rho}{2} {\mid \dot u_{\bf q}\mid}^{2} - \frac{\rho}{2} \cdot
{{\omega}_{q}}^2 {\mid u_{\bf q}\mid}^{2} \right) ,
\label{eq:Sphonon}
\end{equation}
where $ \rho $ is the crystal density and $ u_{\bf -q} = {u_{\bf q}}^*$. Since
we are treating acoustic phonons $ {\omega}_{q} = q \cdot s $, $s$ being the
phonon velocity. ${\cal S}_{int} $ is the action associated with the
interaction between the electron and the phonon environment:

\begin{equation}
{\cal S}_{int} = - \int_{0}^{T}  dt V_{int}\ .
\label{eq:Sinteraction}
\end{equation}
We will treat two electron phonon coupling mechanisms, piezoelectric and
deformation potential, and consider crystals of cubic symmetry. We consider
this symmetry as the most important because most of III-V compounds belong to
this class\cite{re:effective mass}. So for the piezoelectric coupling
\begin{equation}
V_{int}^{piezo} =
\frac{1}{\sqrt{{V}_{vol}}}
\sum_{\bf{q}\nu} \Xi_{\bf{q}\nu} u_{\bf{q}\nu} M_{{\bf q}_{\perp}} \exp (iq_{x} x) ,
\label{eq:mod.piezo}
\end{equation}
where
\begin{equation}
\Xi_{\bf{q}\nu} = {4\pi e\over\epsilon q^{2}} \ \beta_{ijk}q_{i}q_{j}e_{k}^{\nu}
\label{eq:mod.7}
\end{equation}
$\beta_{ijk}$ is the piezoelectric modules, $e$ is the electron charge,
$\epsilon$ is the dielectric constant, $e_{k}^{\nu}$ is the polarization
vector of the $\nu$-th phonon branch, and ${V}_{vol}$ is the normalization
volume. $M_{{\bf q}_{\perp}}$ is the matrix element of the phonon exponent,
$\exp(i{\bf q}_{\perp}{\bf r}_{\perp})$, between the wave functions describing
the electron quantization in the cross section of the wire, ${\bf q}_{\perp}$
and ${\bf r}_{\perp}$ are the phonon wave vector and electron radius vector in
the cross section plane. In cubic crystals the only nonzero components of the
piezoelectric module are $\beta_{xyz}=\beta_{yxz}$ and all other
permutations\cite{landau_ecm}. They are equal and we will designate them as
$\beta$.

The deformation potential couples electrons only with the longitudinal phonon
mode and
\begin{equation}
V_{int}^{def} = \frac{1}{\sqrt{{V}_{vol}}} \sum_{q} i\Lambda \vert q \vert
u_{{\bf q}l} M_{{\bf q}_{\perp}}\exp (iq_{x} x ) \ .
\label{eq:mod.def}
\end{equation}
Here $\Lambda$ is the deformation potential constant.

We neglect the screening of $V_{int}$. Underneath the potential barrier there
are no electrons and the screening by remote electrons is small.

\subsection{Transition amplitude and transition probability}
\label{sb:trp}
We can now define the transition amplitude through the use of the retarded
Green function (\ref{eq:propagator}). $K (x_{b}, u_{{\bf q} f} ,E \mid x_{a},
u_{{\bf q} i})$ is the amplitude to go from initial phonon coordinate $u_{{\bf
q} i}$ to final coordinate $u_{{\bf q} f}$ and from electron coordinate
$x_{a}$ to $x_{b}$ , for a given energy $E$, of the joint system. (In all
intermediate calculation we suppress the index of the phonon branch to
simplify the notations.) The transition amplitude for the joint system to go
from a phonon state designated by $\psi_{0} \left ( u_{{\bf q} i}
\right ) $ to a state $\psi_{n}
\left ( u_{{\bf q} f} \right ) $ is expressed by $K (x_{b}, u_{{\bf q} f} ,E
\mid x_{a}, u_{{\bf q} i})$ in the following manner :

\begin{equation}
\label{eq:ampli}
A_{n b,0 a} = \int du_{{\bf q} i} du_{{\bf q} f}
\psi_{n}^* \left ( u_{{\bf q} f} \right )
K (x_{b}, u_{{\bf q} f} ,E \mid  x_{a}, u_{{\bf q} i})
\psi_{0} \left ( u_{{\bf q} i} \right ) \ .
\end{equation}
We treat the problem with the temperature equal to zero,
therefore phonons are assumed to be initially in the ground state:
$\psi_{0} \left ( u_{{\bf q} i} \right )$.

The transition probability is the absolute value squared of the
transition matrix element. Since we are not interested in the final phonon
configuration, final phonon states are summed over. From the completeness
relation we get the following delta function:
$ \delta\left ( u_{{\bf q} f} - {\tilde u}_{{\bf q} f} \right )$.
We end up with the following expression for the electron transmission
probability:

\begin{eqnarray}
\label{eq:probab}
P_{ b,a} & = & \sum_{n} {\mid A_{n b,0 a} \mid}^2 = \nonumber \\
& = & \int du_{{\bf q} i} \ d{\tilde u}_{{\bf q} i} \ du_{{\bf q} f} \
\psi_{0}^* \left ( {\tilde u}_{{\bf q} f} \right )
K (x_{b}, u_{{\bf q} f} ,E \mid  x_{a} u_{{\bf q} i})
{\tilde K}^* (x_{b}, u_{{\bf q} f} ,E
\mid  x_{a} {\tilde u}_{{\bf q} i})
\psi_{0} \left ( u_{{\bf q} i} \right ) \ .
\end{eqnarray}
This equation defines the transmission coefficient.

\section{Main approximations}
\label{sec:ma}

The usual approach to the calculation of the functional integral in
Eq.(\ref{eq:propagator}) that has been started by Caldeira and
Leggett\cite{Caldeira 81} is the integration out the phonon degrees of
freedom. This can be done exactly because the action is quadratic in $u_{\bf
q}$. In the realistic electron-phonon coupling, this results in a complicated
electron-electron effective potential, since the coupling mechanism is
non-linear in the electron coordinate, thus a simpler approach is needed.

We use the following approach. We consider the situation
(which is typical experimentally) when (i)the barrier is so high that the
tunneling can be considered semi-classically and (ii)the interaction energy
between electrons and phonons is small compared to the height of the barrier.
The first point allows us to make use of the semi-classical approximation to
integrate with respect to all electron paths. The main contribution to this
integral comes from only one optimal trajectory that satisfies the equation
\begin{equation}
m\ddot{x} = - {\partial V_{int}\over\partial x} \ ,
\label{eq:ma.1}
\end{equation}
(for a rectangular barrier $\partial V/\partial x=0$). This equation has to be
solved for given $u(x,t)$. The second point allows us to consider $V_{int}$ in
Eq.(\ref{eq:ma.1}) as a small perturbation and thus it can be neglected. As
usual in tunneling problems,\cite{McLaughlin72,Schulman81} we change the time
$t$ to $-it$ (and $T$ to $iT$) so that the first integral of
Eq.(\ref{eq:ma.1}) has the form
\begin{equation}
{m\dot{x}^{2}\over2} = V - E^{\prime} \ .
\label{eq:ma.2}
\end{equation}
Here $E^{\prime}$ is the integration constant that can be considered as the
electron energy. The value of $E^{\prime}$ is determined from the boundary
conditions $x(0)=x_{a}$ and $x(T)=x_{b}$. For a rectangular barrier when
$x_{a}$ and $x_{b}$ are fixed $E^{\prime}$ is a function of $T$ only. Without
phonon emission $E^{\prime}$ eventually appears equal to the energy of the
incident electron $E$ that is also the total energy. If during tunneling
phonons are emitted then the electron energy $E^{\prime}$ appears smaller than
the total energy.

Thus, in the tunneling problem an important parameter appears the electron
velocity $v=\sqrt{2(V-E^{\prime})/m}$. If $V-E^{\prime}$ is around 10 meV or
larger than in GaAs where $m=0.067m_{0}$ ($m_{0}$ is the free electron mass)
$v$ is around or larger than $2\times10^{7}$ cm/s. This velocity is larger
than the sound velocity $s$ by about two orders of magnitude. That means that
frequencies of phonons with the wave length around the length of the barrier
$L$ are much smaller than the inverse time necessary for an electron to
traverse the barrier. Actually the wave length of a typical phonon interacting
with the tunneling electron is limited not by the length of the barrier but by
the width of the constriction $a$ that is smaller that $L$. However,
practically the ratio $a/L$ for a constriction where tunneling is still
measurable is not very small and the assumption that the typical phonon
frequencies are much smaller than the inverse traverse time is justified. This
assumption means that during the traverse time ($L/v$) the phonon potential
nearly does not change. In the extreme case we can consider it constant. We
call this case the static approximation. A similar approximation was used by
Flynn and Stoneham \cite {Flynn 70} and later by Kagan and Klinger \cite
{Kagan 76} treating the problem of quantum diffusion of atomic particles.

It should be mentioned that the physical significance of the parameter $\omega
t_{0}$, $\omega$ being the phonon frequency, $t_{0}$ the tunneling time as
defined by B\"{u}ttiker and Landauer \cite{Landauer82}, $t_{0}= \frac{L}{v}$
(in our case of a rectangular barrier) was also noted in the two papers of
Ueda and Ando \cite{Ueda 94}, but in their problem the typical frequency
$\omega$ was a tunable parameter defined by the properties of the electric
circuit.

In the static approximation the problem is dramatically simplified because we
need to consider electron motion in a static potential and the stationary
Schr\"odinger equation is enough for this. The integration with respect to
phonon coordinates is reduced to the integration only with respect to $u_{{\bf
q}i}$. The derivation of the corresponding expression for the transmission
coefficient in the static approximation from Eq.(\ref{eq:probab}) with the
help of the expansion in $s/v$ is given in the Appendix.

In the static approximation we can add also $V_{int}^{stat}=V_{int}(u_{{\bf
q}i})$ to the rhs of Eq.(\ref{eq:ma.2}). That immediately shows that the small
parameter characterizing the interaction with phonons is $V_{int}/(V-E)$. In
the calculation of the transmission coefficient we take into account only
terms of the first order in this parameter. So we ignore corrections to the
trajectory $x=vt$ coming from the interaction with phonons. It is obvious in
the calculation of $V_{int}$ and this is also true for ${\cal S}_{el}$.
Indeed, the trajectory $x=vt$ is found from the minimization of ${\cal
S}_{el}$ and any correction to this functional contains a correction to the
trajectory squared. Such an approximation means, in particular, that we
neglect all polaron effects. The physical meaning of the phonon effect in this
approximation is that different phonon configurations change the barrier and
the main contribution to the transmission comes from the configurations
corresponding to the barrier being a bit lower in average, so that the
tunneling probability is higher. It is worthwhile to note that the average
(over configurations) height of the barrier does not change because $\langle
V_{int}\rangle=0$ but nevertheless the average of the tunneling exponent is
modified due to $V_{int}$ (similar to, e.g., $\langle\exp(V_{int})\rangle>1$).

The tunneling across a static barrier is an elastic process and the energy of
the incident electron and the tunneled one is the same. Phonon emission and
the corresponding change of the electron energy come about only in the first
approximation in $s/v$ when the phonon dynamics is taken into account. The
calculation of the dynamic correction to the transmission coefficient that
requires a more complete treatment of the modes presented in the previous
section is given in Sec.\ref{sec:physical results}.

\section {Static approximation}
\label{sec:statapp}

In the static approximation the problem of tunneling can be formulated in a
very simple way, without making use of Eq.(\ref{eq:propagator}). First we can
find the transmission probability for a given static phonon field. The regular
WKB approximation gives for it
\begin{equation}
D(E;u_{{\bf q}i}) = \sqrt{\frac{m}{4
                    \left (
                     V + V_{int}^{stat} - E
                     \right )}}
                     \exp
                \left\{
        - {2\over\hbar} \int_{0}^{L} \sqrt{2m(V + V_{int}^{stat} - E)} \ dx
                \right\} .
\label{eq:sa.1}
\end{equation}
The calculation of the electron transmission coefficient in the case when
phonon field is initially at the ground state is reduced now to the
integration of $D(E;u_{{\bf q}i})$ multiplied by the ground state phonon wave
function squared with respect to $u_{{\bf q}i}$,
\begin{equation}
P(E) = C^{2} \int D(E;u_{{\bf q}i})
        \exp  \left( - \frac{1}{\hbar}
        \sum_{q} \rho q   s {\mid u_{{\bf q}i} \mid}^2
                \right) du_{{\bf q}i}
\label{eq:sa.2}
\end{equation}
(C is a normalizing constant). The expansion in $V_{int}$ gives
\begin{equation}
P(E) = P_{0}(E) C^{2} \int
        \exp    \left\{
        - {\sqrt{2m} \over \hbar\sqrt{V - E}}
        \int_{0}^{L} V_{int}^{stat} dx
        -\frac{ \rho s}{\hbar}
        \sum_{q} q \mid u_{{\bf q} i} \mid^2
                \right\} du_{{\bf q}i} \ ,
\label{eq:sa.3}
\end{equation}
where
\begin{equation}
P_{0}(E) = \frac{\sqrt{m}}{2\sqrt{V - E}} \
           \exp \left\{
        -\frac{2L}{\hbar} \sqrt{2m(V - E)}
                \right\} .
\label{eq:sa.4}
\end{equation}
is the transmission coefficient without interaction with phonons.
Eq.(\ref{eq:sa.3}) can be obtained directly from Eq.(\ref{eq:probab}) (see
Appendix). The result can be written in the form,
\begin{equation}
P(E) = P_{0}(E) \exp \left\{{2\over\hbar} \ \Phi_{stat}(E)\right\} \ .
\label{eq:sa.6}
\end{equation}

For the deformation potential
\begin{equation}
\Phi_{stat}(E) = {\Lambda^{2}L\over8\pi^{2}\rho s_{l}v^{2}}
        \int \vert M_{{\bf q}_{\perp}}\vert^{2} q_{\perp} \
        d^{2}{\bf q}_{\perp}   \ ,
\label{eq:sa.9}
\end{equation}
where $s_{l}$ is the longitudinal phonon velocity.

For the piezoelectric coupling
\begin{equation}
\Phi_{stat}(E) =
        {1\over2\pi^{3}\rho v^{2}}
        \int_{0}^{\infty} dq_{x} {\sin^{2}(q_{x}L/2)\over q_{x}^2}
        \sum_{\nu} {1\over s_{\nu}}
        \int |\Xi_{{\bf q}\nu}|^2\vert M_{{\bf q}_{\perp}}\vert^{2} \
        {d^{2}{\bf q}_{\perp} \over q} \ .
\label{eq:sa.7}
\end{equation}
Because of the anisotropy the result depends on the tunneling direction with
respect to crystalographic axes. For the tunneling in [100] the contribution
of the longitudinal phonons in $\Phi_{stat}$ is small compared to that of
transverse ones in $a/L$ and
\begin{equation}
\Phi_{stat}(E) = {8\beta^{2}e^{2}L \over \rho v^{2}\epsilon^{2}s_{t}}
        \int \vert M_{{\bf q}_{\perp}}\vert^{2}
        {q_{y}^{2}q_{z}^{2} \over q_{\perp}^{5}} \ d^{2}{\bf q}_{\perp} \ ,
\label{eq:sa.8}
\end{equation}
where $s_{t}$ is the velocity of the transverse phonons. For the tunneling in
[110] direction the contributions of both longitudinal and transverse phonons
are of the same order,
\begin{equation}
\Phi_{stat}(E) = {2\beta^{2}e^{2}L \over \rho v^{2}\epsilon^{2}s_{t}}
        \int \vert M_{{\bf q}_{\perp}}\vert^{2}
                \left[
        {q_{y}^{4} + 4q_{y}^{2}q_{z}^{2} \over s_{t}q_{\perp}^{4}} -
        \left({1 \over s_{t}} - {1 \over s_{l}}\right)
        {9q_{y}^{4}q_{z}^{2} \over q_{\perp}^{6}}
                \right]
        {d^{2}{\bf q}_{\perp} \over q_{\perp}} \ ,
\label{eq:sa.10}
\end{equation}

As can be seen the typical phonon wave numbers with which the electron
interacts are fixed through the length scales of the problem, the barrier
length $L$ and the width of the constriction $a$. It is important to note
that, the static correction, in the exponent is positive, therefore static
phonons reduce the effective potential barrier height, enhancing electron
tunneling probability. The electron, due to coupling to the zero point
fluctuations of the phonon field, tunnels through an effectively lower
potential barrier.

\section {dynamic corrections and energy loss}
\label{sec:physical results}

To get the electron's energy loss due to phonon emission during the tunneling
process, one needs to go beyond the static approximation. As the first step we
make use of the WKB approximation to simplify the expression for the
propagator $K( x_{b},u_{{\bf q} f},T\mid x_{a},u_{{\bf q} i})$
(\ref{eq:propagator}). In this approximation the main contribution to the
integral with respect to $x(t)$ comes from the saddle point trajectory that in
the first approximation in $V_{int}/(V-E)$ is determined by
Eq.(\ref{eq:ma.2}), As a result the propagator is factorized.
In Eq.(\ref{eq:ma.2}) we passed to the imaginary time so
the trajectory is $x_{0}(t)=vt$ where $v=\sqrt{2(V-E^{\prime})/m}$, thus
the propagator can be expressed as:

\begin{eqnarray}
K (x_{b}, u_{{\bf q} f} ,-iT \mid x_{a}, u_{{\bf q} i}) =
        K_{0}(x_{b},T \mid x_{a})
        K_{ph}(u_{{\bf q} f},T \mid u_{{\bf q}i}) \ .
\label{eq:dc.1}
\end{eqnarray}
Here
\begin{equation}
K_{0}(x_{b},T \mid x_{a}) =
                \left[
        {m \over 4(V + V_{int}^{stat} - E)}
                \right]^{1/4}
        \exp    \left ( \frac{- E'T}{\hbar} \right )
        \exp{ \left (- \frac{1}{\hbar} \int_{0}^{L} dx
        \sqrt{2 m (V - E')} \right )}
\label{eq:dc.2}
\end{equation}
is the electron propagator without interaction with phonons and
\begin{equation}
K_{ph}(u_{{\bf q}f},T \mid u_{{\bf q}i}) =
        \int {\cal D} u_{\bf q}
        \exp    \left\{
        - {1\over\hbar}
                \left[
        {\cal S}_{ph}(u_{\bf q}) + {\cal S}_{int}(u_{\bf q}, x_{0})
                \right]
                \right\} ,
\label{eq:dc.3}
\end{equation}
is the phonon part of the propagator.

So as the electron trajectory is determined the phonon part of the propagator
is the propagator of an ensemble of forced harmonic oscillators.\cite{Hibbs65}
The integration with respect to $u_{\bf q}$ leads to
\begin{equation}
K_{ph}(u_{{\bf q}f}, T \mid u_{{\bf q}i}) =
        g(T) \exp       \left[
        - {1\over\hbar} \ {\cal S}_{cl}(u_{{\bf q}f}, T \mid u_{{\bf q}i})
                        \right] ,
\label{eq:dc.4}
\end{equation}
where
\begin{equation}
g(T) = \prod_{\bf q} \sqrt{\rho\omega_{q} \over 2\pi\hbar\sinh\omega_{q}T} \ ,
\label{eq:dc.5}
\end{equation}
and
\begin{eqnarray}
{\cal S}_{cl}(u_{{\bf q}f}, T \mid u_{{\bf q}i}) & = &
        \sum_{q} \frac{\rho \omega_{q}}{2 \sinh(\omega_{q} T)}
                \Big\{
\cosh (\omega_{q} T) \left ( {\mid u_{{\bf q} i} \mid}^2 +
{\mid u_{{\bf q} f} \mid}^2 \right ) - \left (
 u_{{\bf q} i}{u_{{\bf q} f}^*} + {u_{{\bf q} i}^*}u_{{\bf q} f} \right )
\nonumber \\ && \hspace{-2cm} -
        u_{{\bf q} f}
                \left(
        \frac{1}{\rho\omega_{q} v}
        \int_{0}^{L} dx f_{\bf q}(x) \sinh(\frac{\omega_{q} x}{v})
                \right ) -
        u_{{\bf q} i}
                \left(
        \frac{1}{\rho\omega_{q} v}
        \int_{0}^{L} dx f_{\bf q}(x) \sinh(\frac{\omega_{q} (L-x)}{v})
                \right)
\nonumber\\ && \hspace{-2cm} -
        {u_{{\bf q} f}^*}
                \left(
        \frac{1}{\rho\omega_{q} v}
        \int_{0}^{L} dt f_{\bf q}^{*}(x) \sinh(\frac{\omega_{q} x}{v})
                \right)
        - {u_{{\bf q} i}^*}
                \left(
        \frac{1}{\rho\omega_{q} v}
        \int_{0}^{L} dx f_{\bf q}^*(x) \sinh (\frac{\omega_{q} (L-x)}{v})
                \right)
\nonumber\\ && \hspace{-2cm} -
        \frac{2}{{\rho}^2{\omega_{q}}^2 {v}^2}
        \int_{0}^{L} dx \int_{0}^{x} dy
        \left[ f_{\bf q}(x)f_{\bf q}^*(y) + f_{\bf q}^*(x)f_{\bf q}(y) \right]
        \sinh (\frac{\omega_{q} y}{v}) \sinh(\frac{\omega (L-x)}{v})
                \Big\} \ .
\label{eq:dc.6}
\end{eqnarray}
Here according to Eqs.(\ref{eq:mod.def}) and (\ref{eq:mod.piezo})
\begin{equation}
f_{\bf q}(x) = {\Xi M_{{\bf q}_{\perp}} \over \sqrt{V_{vol}}} \ e^{iq_{x}x}
\label{eq:dc.7}
\end{equation}
for the piezoelectric interaction and
\begin{equation}
f_{\bf q}(x) =i \vert q \vert {\Lambda M_{{\bf q}_{\perp}}\over \sqrt{V_{vol}}}
              \ e^{iq_{x}x}
\label{eq:dc.8}
\end{equation}
for the deformation potential interaction. One should note that in
Eq.(\ref{eq:dc.6}) we made a transformation of variables from $t$ in the
action, to $x=vt$.

Making use of the factorization (\ref{eq:dc.1}) the expression for tunneling
probability (\ref{eq:probab}) can be written as
\begin{equation}
\label{eq:tranprob}
P(E) = \int dT \int d\tilde{T}
        K_{0}(x_{b},T \mid x_{a})K_{0}(x_{b},\tilde{T}
        \mid x_{a}) \xi_{ph} \left( T ; \tilde{T} \right) ,
\end{equation}
where
\begin{eqnarray}
\label{eq:dc.10}
& & \xi_{ph}\left( T ; \tilde{T}\right) = g(T)g(\tilde{T})
\nonumber \\
&\times&\int du_{{\bf q}i} \int du_{{\bf q}f} \int {d\tilde u}_{{\bf q}i}
        \exp    \left[
        - {1\over\hbar} \
        {\cal S}_{cl}\left( u_{{\bf q}f},T \mid u_{{\bf q} i} \right)
        - {1\over\hbar} \
        {\cal S}_{cl}\left(u_{{\bf q}f},\tilde{T}, \mid {\tilde u}_{{\bf q}i}\right)
                \right]
        {\psi_{0} \left ( u_{{\bf q} i} \right )}
        {\psi_{0}^* \left ( {\tilde u}_{{\bf q} i}\right )} \ ,
\end{eqnarray}
and $\psi_{0}\left(u_{{\bf q} i}\right)$ is the phonon ground state wave
function. The integrals with respect to $T$ and $\tilde{T}$ in
Eq.(\ref{eq:tranprob}) are calculated by the saddle point
method\cite{McLaughlin72}. Due to the symmetry of
$\xi_{ph}\left[T;\tilde{T}\right]$ with respect to the transposition of $T$
and $\tilde{T}$ the saddle point values of these variables are identical. We
are interested only in the exponential part of the transition probability and
for this only $\xi_{ph}\left[T;T\right]$ is necessary. The result can be
written in the form
\begin{equation}
\xi_{ph}\left(T;T\right) = A \exp\left[ {2\over\hbar} \ \Phi_{ph}(T)\right] \ ,
\label{eq:dc.11}
\end{equation}
where $\Phi_{ph}(T)$ is calculated in the Appendix and the pre-exponential
factor $A$, will not be calculated.

We now break $\Phi_{ph}(T)$ into a static part, $\Phi_{stat}$, and a dynamical
correction so that $\Phi_{ph}=\Phi_{stat}+\Phi_{dyn}$. As one can expect the
static part is identical to $\Phi_{stat}(E^{\prime})$ obtained in
Sec.\ref{sec:statapp}, Eq.(\ref{eq:sa.8}), in a more simple way. The dynamical
correction, $\Phi_{dyn}(T,E^{\prime})$, is obtained in the Appendix in the
leading order in $s/v\ll1$. Using these notations the saddle point equation
for time integration is given by:
\begin{equation}
\label{eq:tisp}
                \left(
        {\partial\Phi_{el}\over\partial E'} -
        {\partial\Phi_{stat}\over\partial E'} - T
                \right)
        \frac{dE'}{dT} - \frac{d \Phi_{dyn}}{dT}
        + E - E' = 0 \ ,
\end{equation}
where $\Phi_{el}(E^{\prime})=L\sqrt{2m(V-E^{\prime})}$.

According to Sec.\ref{sec:statapp},
$\Phi_{el}(E^{\prime})-\Phi_{stat}(E^{\prime})$ is the electron action in the
phonon static field at the trajectory with the energy $E^{\prime}$. The
derivative of the action with respect to the energy is the traveling time
along the trajectory. The first term, $\partial\Phi_{el}/\partial E'$, has
been defined by B\"uttiker and Landauer\cite{Landauer82} as a semiclassical
traverse time. The second term, $\partial\Phi_{stat}/\partial E'$, gives a
phonon correction to the traverse time. The sum to the two terms equals $T$
and the expression in the parentheses is identically zero. Then
Eq.(\ref{eq:tisp}) can be written in the form
\begin{equation}
\label{eq:sadlepoint}
E - E' = - \frac{\partial \Phi_{dyn}}{\partial v} \frac{v^2}{L} .
\end{equation}
In the accepted approximation in the rhs of this equation the difference
between $E$ and $E^{\prime}$ has to be neglected.

According to our definition the energy $E$ appearing at the Fourier transform
of the transmission amplitude is the total energy of the system while
$E^{\prime}$ is the energy characterizing the electron trajectory. The
difference between them is the energy transferred to the phonon system, i.e.,
the average energy loss of the tunneling electron.

The substitution of $T=L/v$ and calculation of the integral with respect to
$q_{x}$ in Eq.(\ref{eq:ap.7}) we obtain for the deformation potential,
\begin{equation}
\Phi_{dyn}(E) =
        {{\Lambda}^2 L^{2}\over 8\pi^{2}\rho v^{3}}
        \int \vert M_{{\bf q}_{\perp}}\vert^{2} \ {q_{\perp}}^2
        {d^{2}{\bf q}_{\perp}} \ .
\label{eq:dc.12}
\end{equation}
For the piezoelectric potential the terms in $|\Xi_{{\bf q}\nu}|^2$ containing
$q_{x}$ are small in $a/L$ and the main contribution is
\begin{equation}
\Phi_{dyn}(E) =
        {L^{2}\over 8\pi^{2}\rho v^{3}} \sum_{\nu}
        \int |\Xi_{{\bf q}\nu}|^2 \vert M_{{\bf q}_{\perp}}\vert^{2} \
        {d^{2}{\bf q}_{\perp}} \ .
\label{eq:dc.13}
\end{equation}
For [100] tunneling direction Eq.(\ref{eq:dc.13}) becomes
\begin{equation}
\Phi_{dyn}(E) =
        {8\beta^{2}e^{2}L^{2}\over \rho v^{3}\epsilon^{2}}
        \int \vert M_{{\bf q}_{\perp}}\vert^{2} \
        {q_{y}^{2}q_{z}^{2} \over q_{\perp}^{4}} \
        {d^{2}{\bf q}_{\perp}} \ ,
\label{eq:dc.14}
\end{equation}
and for [110] tunneling direction
\begin{equation}
\Phi_{dyn}(E) =
        {2\beta^{2}e^{2}L^{2}\over \rho v^{3}\epsilon^{2}}
        \int \vert M_{{\bf q}_{\perp}}\vert^{2} \
        (q_{y}^{4} + 4q_{y}^{2}q_{z}^{2}) \
        {d^{2}{\bf q}_{\perp} \over q_{\perp}^{4}} \ .
\label{eq:dc.15}
\end{equation}

It worthwhile to note that $\Phi_{dyn}$ is positive, i.e., it increases the
transmission coefficient as well as $\Phi_{stat}$. One could expect that
phonon emission makes the barrier for electron effectively higher that may
lead to a reduction of the transmission probability. However, $\Phi_{dyn}$ is
calculated for $E=E^{\prime}$ and this effect can appear only in the next
approximation.

The average energy loss resulting from the deformation potential coupling is,
\begin{equation}
E - E^{\prime} =
        {3{\Lambda}^2 L\over 8\pi^{2}\rho v^{2}}
        \int \vert M_{{\bf q}_{\perp}}\vert^{2}{q_{\perp}}^2 \
        {d^{2}{\bf q}_{\perp}} \ .
\label{eq:dc.16}
\end{equation}
Whereas the average energy loss due to the piezoelectric coupling is given by:
\begin{equation}
E - E^{\prime} =
        {24\beta^{2}e^{2} L\over \rho v^{2}\epsilon^{2}}
        \int \vert M_{{\bf q}_{\perp}}\vert^{2} \
        {q_{y}^{2}q_{z}^{2} \over q_{\perp}^{4}} \
        {d^{2}{\bf q}_{\perp}} \ ,
\label{eq:dc.17}
\end{equation}
for [100] tunneling direction and
\begin{equation}
E - E^{\prime} =
        {6\beta^{2}e^{2} L\over \rho v^{2}\epsilon^{2}}
        \int \vert M_{{\bf q}_{\perp}}\vert^{2} \
        (q_{y}^{4} + 4q_{y}^{2}q_{z}^{2}) \
        {d^{2}{\bf q}_{\perp} \over q_{\perp}^{4}} \ .
\label{eq:dc.18}
\end{equation}
for [110] tunneling direction.

 The energy loss is proportional to the width of the barrier which means
that it is accumulated along it.

The comparison the static and dynamical phonon corrections to the tunneling
probability gives
        $\Phi_{dyn}/\Phi_{stat}\sim(s/v)(L/a)$ where $a$ is the width of the
constriction. Typically the ratio $L/a$ is not very large so the expansion
that we used is justified.

\section{discussion and summary}
\label{sec:discussion}

In this paper we have presented a detailed study of effect of coupling with
acoustic phonons on electron tunneling across a rectangular barrier in a
realistic situation. We studied piezoelectric and deformation potential
coupling at low temperature which means that $\lambda_{T}/a\gg1$, where
$\lambda_{T}$ is the thermal phonon wave length (see Appendix). We considered
only a one-dimensional problem that can be realized in a quantum wire or a
narrow constriction.

In our calculation we assumed that the barrier is high enough to describe
tunneling in the semiclassical approximation. Our detailed calculations reveal
that the typical phonon interacting with the tunneling electron is chosen
through the length of the barrier $L$ and the width of the constriction or
quantum wire $a$. It is thus the geometry of the potential barrier which
defines the phonon frequency $\omega$. Under such a condition the electron
motion under the barrier is so fast that phonons don't follow it and can be
considered nearly static during the time necessary for electron to traverse
the barrier. The main effect of phonons in this case is a modulation of the
barrier so that its height has to be considered as a quantum mechanical
variable which probability distribution is controlled by a phonon wave
function. In this case, roughly speaking, an electron chooses for tunneling
the phonon configurations when the barrier is lower than its average value,
that results in an increase of the transmission coefficient compared to that
with zero electron-phonon coupling. Thus the interaction of an electron with
the zero point phonon fluctuations increases the tunneling probability.

The correction to the tunneling exponent in the static approximation is
proportional to the length of the barrier. That is the coupling affects the
dependence of the transmission coefficient on the height of the barrier only.

The dynamical correction leads to two effects. First, it describes the
reduction of the electron energy due to phonon emission. Second, which is
probably more interesting, it gives the dependence of the transmission
coefficient on the length of the barrier different from the regular one where
the log of the transmission coefficient is linear in the length. We calculated
only the first order correction to the exponential dependence. The effect can
be stronger and is probably measurable for tunneling near the top of the
barrier. One should note that even though the dynamical correction to the
transition probability is smaller than the static correction, it is still an
exponential correction and it easily can be made larger than unity, i.e., by
changing the length of the barrier. In this case it can significantly affect
the tunneling probability.

The dependence of transmission coefficient on the height and the length of a
barrier can be measured in devices where the barrier is introduced with the
help of a gate. In such a device both the height and the length of the barrier
are controlled by the gate voltage. Our results point out to a deviation of
these dependencies from the standard ones obtained from stationary
Schr\"odinger equation.

The application of our results to a long quantum wire can encounter a
difficulty because we did not take into account electron-electron interaction.
The effect of this interaction can be reduced for a short wire or a narrow
constriction.

We expect non-trivial results through further use of the piezoelectric and
deformation potential coupling mechanisms in two or three dimensional systems.

We believe that the physical considerations presented here which
greatly simplified the calculations are convenient for extending the
calculations to two and three dimensional physical situations.

\newpage

\appendix

\section*
{non-static calculation of the transition amplitude and transition
probability}

In the Appendix we carry out the integration with respect to initial and final
phonon coordinates and expand the result in two small parameters defined in
Sec.\ref{sec:ma}, $s/v$ and $V_{int}/(V-E)$. The calculation presented here
are for the finite temperature case.

The explicit form of Eq.(\ref{eq:dc.10}) for $\tilde{T}=T$ is
\begin{eqnarray}
\label{eq:ap.1}
& & \xi_{ph}\left( T ; T\right) = g^{2}(T)
\nonumber \\
&\times&\int du_{{\bf q}i} \int du_{{\bf q}f} \int {d\tilde u}_{{\bf q}i}
        \exp    \left[
        - {1\over\hbar} \ \sum_{\bf q} \alpha_{q}(T)
       \eta_{q}\left(u_{{\bf q}f},u_{{\bf q} i},{\tilde u}_{{\bf q}i};T \right)
                \right]
         \frac{1}{Z} \sum_{m} e^{-\beta E_{m}}
        {\psi_{m} \left ( u_{{\bf q} i} \right )}
        {\psi_{m}^* \left ( {\tilde u}_{{\bf q} i}\right )} \ ,
\end{eqnarray}
where $Z=\sum_{m} e^{-\beta E_{m}}$ , $\beta$ is one over the Boltzmann
constant times the temperature (we use this notation only in the Appendix and
it should not be confused with the piezoelectric module in the body of the
paper),
\begin{eqnarray}
\eta_{q}\left(u_{{\bf q}f},u_{{\bf q} i},{\tilde u}_{{\bf q}i};T \right)
        & = & \epsilon_{q} \left ( {\mid u_{{\bf q} i} \mid}^2 +
{\mid u_{{\bf q} f} \mid}^2 \right ) -\left (
u_{{\bf q} i}{u_{{\bf q} f}^*} + {u_{{\bf q} i}^*}u_{{\bf q} f} \right )
\nonumber \\
& + & \epsilon_{q}
\left ( {\mid \tilde{u_{{\bf q} i}} \mid}^2 +{\mid u_{{\bf q} f} \mid}^2
\right ) - \left (\tilde{u_{{\bf q} i}}{u_{{\bf q} f}^*}
+\tilde{ {u_{{\bf q} i}^*}}u_{{\bf q} f} \right ) \nonumber \\
& + & u_{{\bf q} i}I_{2} + u_{{\bf q} f}I_{1} + u_{{\bf q} i}^*I_{2}^* +
u_{{\bf q} f}^*I_{1}^* \nonumber \\
& + & \tilde{u_{{\bf q} i}^*}I_{2}^* + u_{{\bf q} f}^*I_{1}^* +
\tilde{u_{{\bf q} i}}I_{2} + u_{{\bf q} f}I_{1} - I_{3} \ ,
\label{eq:ap.2}
\end{eqnarray}
$\alpha_{q}=\rho\omega_{q}/2\sinh(\omega_{q} T)$,
$\epsilon_{q} =\cosh(\omega_{q}T)$,
\begin{mathletters}
\begin{eqnarray}
I_{1} & = & \frac{1}{\rho\omega_{q} v}
        \int_{0}^{L} dx f_{\bf q}(x) \sinh( \frac{\omega_{q} x}{v} ) \ ,
\label{eq:ap.3a} \\
I_{2} & = & \frac{1}{\rho\omega_{q} v}
        \int_{0}^{L} dx f_{\bf q}(x) \sinh (\frac{\omega_{q} (L-x)}{v} ) \ ,
\label{eq:ap.3b} \\
I_{3} & = & \frac{2}{{\rho}^2{\omega_{q}}^2 {v}^2}
        \int_{0}^{L} dx \int_{0}^{x}dy
                \left[
        f_{\bf q}(x)f_{\bf q}^*(y) + f_{\bf q}^*(x)f_{\bf q}(y)
                \right]
        \sinh (\frac{\omega_{q} y}{v})
        \sinh ( \frac{\omega_{q} (L-x)}{v} ) \ .
\label{eq:ap.3c}
\end{eqnarray}
\label{eq:ap.4}
\end{mathletters}
Using the well known result for the density matrix \cite{Feynman 72}:
\begin {equation}
\label{eq:ap.4a}
\frac{1}{Z} \sum_{m} e^{-\beta E_{m}}
        {\psi_{m} \left ( u_{{\bf q} i} \right )}
         {\psi_{m}^* \left ( {\tilde u}_{{\bf q} i}\right )}=
         g(\beta)\exp \left [ - \frac{2}{\hbar} \sum_{\bf q} \rho\omega_{q}
         {\mid \tilde{u_{{\bf q} i}} \mid}^2 \tanh (\frac{\hbar \omega_{q} \beta}
 {2} ) \right ]
\end{equation}
where $g(\beta)$ is given by:
\begin{equation}
g(\beta) = \prod_{\bf q} \sqrt{\rho\omega_{q} \over 2\pi\hbar\sinh \hbar \omega_{q}
\beta} .
\label{eq:ap.4b}
\end{equation}

The result of the integration with respect to initial and final phonon
coordinates gives $\xi_{ph}\left(T;T\right)=A\exp[2\Phi_{ph}(T)/\hbar]$ where
\begin{equation}
\Phi_{ph}(T) =  \sum_{q} \frac{\alpha_{q}}{\epsilon_{q}}
                \left[
        {\mid \epsilon_{q} \cdot I_{2} + I_{1} \mid^2 \over
        {\epsilon_{q}}^2 + \rho s q \gamma_{q}/2 - 1} +
        {\mid I_{1} \mid}^2 + \frac{\epsilon_{q} \cdot I_{3}}{2}
                \right] ,
\label{eq:ap.5}
\end{equation}
{\bf $ \gamma_{q}= \epsilon_{q} \tanh( \frac{\hbar \omega_{q} \beta}
{2}) / \alpha_{q}$}

Finite temperature effects in Eq.(\ref{eq:ap.5}) can be neglected if
$\tanh(\hbar\omega_{q}\beta/2)\approx1$. The physical meaning of this
approximation is that the phonon thermal wave length is much larger than the
quantum wires width i.e. $\lambda_{T}/a\gg 1$. For the width of the
constriction of 100 \AA~ or larger this condition is satisfied up to room
temperature. From this estimate it is also follows that the matrix element for
high energy phonons that could activate a tunneling electron to energy
comparable with the height of the barrier is small and we neglect it. Thus the
rest of the calculations will be performed for the zero temperature case.

One should note that the phonon part of the transition probability,
$\xi_{ph}(T;T)$, is larger than 1, causing an enhancement of the electron
tunneling probability.

It is convenient to separate in $\Phi_{ph}(T)$ a static part that is obtained
by putting $s/v=0$ and a dynamical correction,
\begin{equation}
\Phi_{ph}(T) = \Phi_{stat}(T) + \Phi_{dyn}(T) \ .
\label{eq:ap.6}
\end{equation}
The static part
\begin{eqnarray}
\label{eq:ap.7}
\Phi_{stat} = \sum_{q} \frac{1}{\rho q s v^{2}}
        \left\vert \int_{0}^{L} dx f_{\bf q}(x) \right\vert^2 ,
\end{eqnarray}
is equal to the expression appearing in Eq.(\ref{eq:sa.7}). The dynamical
correction is necessary for the calculation of phonon emission and it is
calculated in the leading order in $s/v\ll1$. For the deformation potential
the dynamical correction is given by:
\begin{eqnarray}
\Phi_{dyn}(E) & = &
        {4\Lambda^{2}T \over \rho v^{2}V_{vol}}
        \sum_{\bf q}
       \left [ {\sin(q_{x}L) \over q_{x}^{3}L} - {\cos(q_{x}L) \over q_{x}^{2}}
        \right]
       {q_{\perp}}^2  \vert M_{{\bf q}_{\perp}}\vert^{2} \ .
\label{eq:ap.9}
\end{eqnarray}

For the piezoelectric coupling one gets:
\begin{equation}
\Phi_{dyn}(E) =
        {4T \over \rho v^{2}V_{vol}}
        \sum_{{\bf q}\nu}
                \left[
        {\sin(q_{x}L) \over q_{x}^{3}L} - {\cos(q_{x}L) \over q_{x}^{2}}
                \right]
        |\Xi_{{\bf q}\nu}|^{2}\vert M_{{\bf q}_{\perp}}\vert^{2} \ .
\label{eq:ap.8}
\end{equation}


\begin{thebibliography}{99}
\bibitem{re:effective mass}
Though we are treating a general case not confined to a certain material, the
sort of physical arrangement we are treating, quantum wires and narrow
constrictions are currently fabricated mainly with the use of GaAs technology.
\end{thebibliography}
\end{document}